\documentclass[reprint,a4paper,twoside,nofootinbib,showkeys,reprint,twocolumn,superscriptaddress,floatfix]{revtex4-1}
\usepackage[english]{babel}
\usepackage[utf8]{inputenc}
\usepackage{graphicx}
\usepackage{mathtools}
\usepackage{placeins}
\usepackage[T1]{fontenc}
\usepackage{lipsum}
\usepackage{csquotes}
\usepackage{siunitx}
\usepackage[export]{adjustbox} 
\usepackage{amsmath}
\usepackage{amssymb}
\usepackage{bm}
\usepackage{ragged2e}
\usepackage{xcolor}
\usepackage{float}
\usepackage[pdftex,colorlinks=true, linkcolor=blue, citecolor=blue]{hyperref}
\newcommand{\ba}{\begin{eqnarray}}
\newcommand{\ea}{\end{eqnarray}}
\newcommand{\be}{\begin{equation}}
\newcommand{\ee}{\end{equation}}
\newcommand{\bi}{\begin{itemize}}
\newcommand{\ei}{\end{itemize}}
\newcommand{\bra}[1]{\langle #1 |}
\newcommand{\ket}[1]{| #1 \rangle}
\newcommand{\ds}{\displaystyle}
\newcommand{\MPl}{M_{\mathrm{Pl}}}

\newcommand{\eqequiv}{\hspace{-1mm} & \equiv & \hspace{-1mm}}

\newcommand{\LF}{\left(}
\newcommand{\RF}{\right)}
\newcommand{\LT}{\left[}
\newcommand{\RT}{\right]}
\newcommand{\Ld}{\left.}
\newcommand{\Rd}{\right.}

\newcommand{\pha}{\widehat{p}_{A}}
\newcommand{\phb}{\widehat{p}_{B}}
\newcommand{\xha}{\widehat{x}_{A}}
\newcommand{\xhb}{\widehat{x}_{B}}

\newcommand{\2}{\frac{1}{2}}

\newcommand{\tth}{\frac{2}{3}}

\newcommand{\mr}{\mathrm}

\newcommand{\mP}{\mr{P}}
\newcommand{\mD}{\mr{D}}

\newcommand{\hs}{\hspace{5mm}}
\newcommand{\vs}{\vspace{0mm}\\}
\newcommand{\non}{\nonumber\\}

\labelformat{equation}{Eq.~(#1)} 
\labelformat{appendix}{Appendix #1}

\newcommand{\Hc}{\mathcal{H}}

\begin{document}

\title{ Probing massless and massive gravitons via entanglement in a warped extra dimension }
\author{Shafaq Gulzar Elahi}
\email{ug18sge@iacs.res.in}
\affiliation{Indian Association for the Cultivation of Science, Kolkata, India}

\author{Anupam Mazumdar}
\affiliation{Van Swinderen Institute,Nijenborgh 4, 9747 AG, University of Groningen, The Netherlands}

\begin{abstract}
Gravity's quantum nature can be probed in a laboratory by witnessing the entanglement between the two quantum systems, which cannot be possible if gravity is a classical entity. In this paper, we will provide a simple example where we can probe the effects of higher dimensions, in particular the warped extra dimension of five-dimensional Anti-de Sitter spacetime ($\rm AdS_5$). We assume that the two quantum harmonic oscillators are kept at a distance $d$ on a 3-brane (our 4D world) embedded in $\rm AdS_5$, while gravity can propagate in all five dimensions. We will compute the effective potential due to the massless and  massive gravitons propagating in the warped geometry.  We will compute the entanglement between position and momentum states for both static and non-static cases. The entanglement enhances compared to the four-dimensional massless graviton, and it depends now on the $\rm AdS_5$ radius. We will also show that if we would prepare non-Gaussian superposition states, e.g. spatial superposition of masses of order $10^{-14}-10^{-15}$kg with a superposition size of ${\cal O}(20)$ micron, we can yield larger concurrence of order ${\cal O}(0.1)$.

\end{abstract}
\maketitle

\section{Introduction}

Entanglement is a unique quantum feature that cannot be mimicked by any classical theory~\cite{Horodecki}. By witnessing the entanglement between the two quantum objects, authors of Ref.~\cite{Bose:2017nin} proposed to probe the quantum nature of gravity in a laboratory~\footnote{ First reported the results and scheme of Ref.~\cite{Bose:2017nin} in a talk~\cite{ICTS}.}, see also~\cite{Marletto}. The experiment  proposed by the authors in \cite{Bose:2017nin} is known as the QGEM (quantum gravity-induced entanglement of masses) protocol, where the idea is to probe the quantum nature of gravity via spin entanglement. Recently, also a new protocol has been presented to test the spin-2 nature of gravity in an entanglement test between a quantum matter and a laser beam in a cavity~\cite{Biswas:2022qto}. All these protocols rely on a powerful theorem, known as the LOCC theorem, where LOCC stands for local operation and classical communication~\cite{Bennett}. The LOCC theorem suggests that if the two quantum systems are not entangled, to begin with then they will remain unentangled if the interaction between the two quantum systems remains classical in nature. Similarly, if the gravitational interaction is classical then the two quantum systems will never get entangled~\cite{Marshman:2019sne, Carney,Stamp,Miao}. In a perturbative quantum gravity, one can show this explicitly in a canonical approach~\cite{QGEM}, path integral approach~\cite{Chistodoulou}, and a very potent tool of axiomatic quantum gravity~\cite{Wald}. The QGEM protocol utilizes the scheme where the two masses are kept at a distance in a quantum spatial superposition, e.g. Schr\"odinger Cat state, for a time $\tau$. The only allowed interaction is assumed to be solely gravity, it is possible to mitigate electromagnetic interactions, such as Coulomb, dipole-dipole, and higher-order interactions, still, they remain the biggest challenge to mitigate in an experimental setup ~\cite{Bose:2017nin,Vankamp}. Of course, there are many challenges, such as creating massive quantum superposition~\cite{Scala,Wan,Pedernales,Wood,Marshman:2021wyk,Zhou:2022frl,Zhou:2022jug,Zhou:2022epb}, keeping the system intact from various sources of decoherence~\cite{Bassi,Hornberger,Isart,Tilly:2021qef,Schut:2021svd,Rijavec,Gunnick}, and above all protecting the system from gravity gradient~\cite{Toros:2020dbf,Wu:2022rdv}, and relative acceleration noise~\cite{Toros:2020dbf}.

Despite these challenges, the QGEM experiment is feasible and it tests the nature of quantum gravity in a similar spirit as Bell's test~\cite{Hensen}. The crucial observation is that the quantum correlation/entanglement exists despite $\hbar \rightarrow 0$, as was first illustrated in the two quantum systems with a large angular momentum~\cite{Peres,Gisin}. Although extracting the entanglement will become extremely challenging. In a similar vein, the QGEM protocol can test the quantum nature of gravity at the lowest order from the Newtonian potential. 
\begin{figure}[ht]
    \centering
   { \includegraphics[width=150pt, max width=\textwidth]{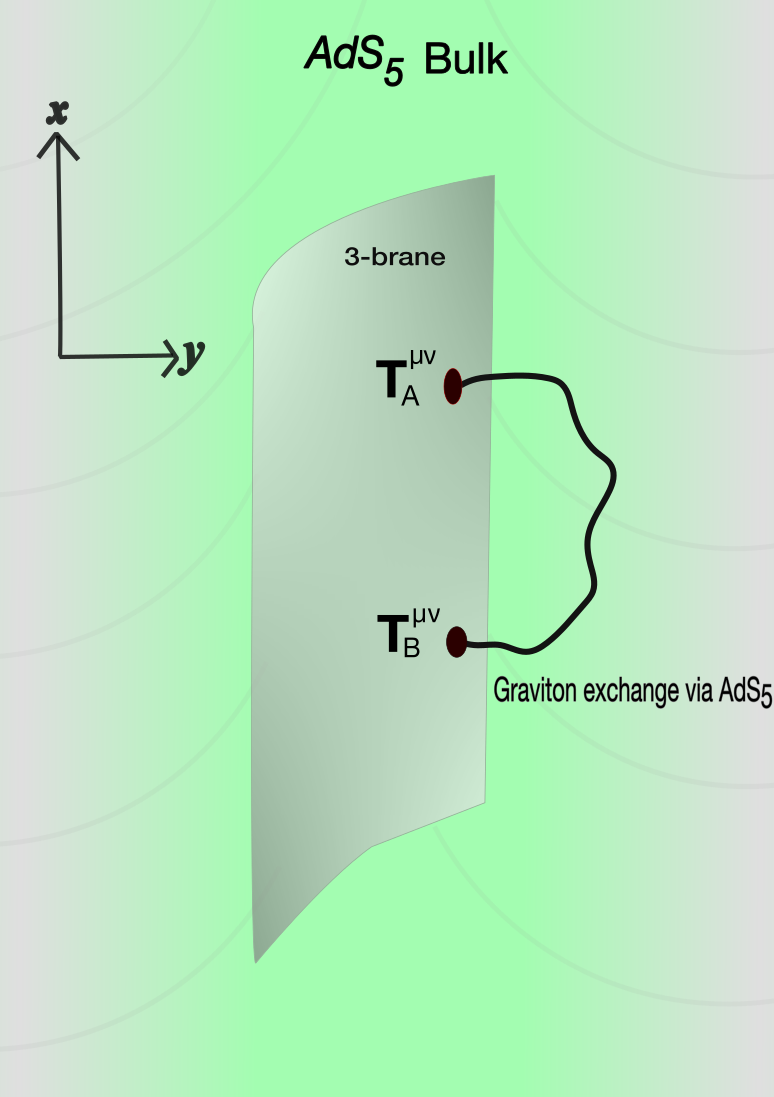}}
    \caption{Two particles on the 3-brane interacting via graviton exchange from the bulk.}
    \label{fig:my_label}
\end{figure}
\FloatBarrier
In an effective field theory approach to quantum gravity, the quantization of spin-2 graviton can be shown to yield the Newtonian potential, the light bending effect due to the gravitational potential, and higher order contributions including the effects of rotation via computing the scattering amplitude and taking appropriately the non-relativistic limit. The quantization of graviton can be followed either via Gupta's approach or via path integral approach by imposing gauge fixing contribution~\cite{QGEM,Chistodoulou}. Although the experiment will never probe graviton directly, very indirectly we will probe the graviton's properties that will manifest the entanglement, and the equivalence principle~\cite{Bose:2022czr}.

In this paper, we wish to probe how the entanglement is developed if we depart from 4-D Minkowski spacetime and assume that there exists a warped extra spatial dimension, e.g. in five-dimensional (5-D) Anti-de Sitter ($\rm AdS_5$) spacetime. The search for extra dimensions has been an active area of research ever since Kaluza-Klein proposed a geometric interpretation of electromagnetism through the introduction of an extra spatial dimension and this interest was revived with the development of string theory which predicts that spacetime is fundamentally higher dimensional. Ever since this concept was introduced, it has been studied to explain various pressing issues in particle physics, ranging from issues related to  the resolution of  gauge hierarchy/fine tuning  problem~\cite{fine:tuning}, the origin of neutrino masses~\cite{neutrino}-\cite{neutrino2}, fermion mass hierarchy\cite{fermion} and dark matter  to large scale phenomena such as inflation\cite{inflation,inflation2}, bouncing~\cite{bounce}-\cite{bounce3} phenomena in cosmology as well  galactic structure\cite{galaxy}-\cite{galaxy3} in astrophysics.  

Two extra-dimensional  models namely large extra dimensions~\cite{large}-\cite{large3},\cite{ADD} and warped extra dimensions\cite{warp}-\cite{warp3},\cite{RandallSundrum1,RandallSundrum2}  became extremely popular at the beginning of this century. The testing beds of these models ranged from collider physics to cosmological/astrophysical scenarios. 
In the context of the fine-tuning problem related to the  large radiative corrections to Higgs mass, the warped geometry model by Randall and Sundrum~\cite{RandallSundrum1}  turned out to be very  successful as it could resolve the problem without introducing any intermediate scale in theory. Interestingly String theory can provide an analog of such a warped extra-dimensional scenario through a  throat-like  geometry~\cite{KS} and 
Randall Sundrum model can capture the essential features of this throat geometry in a simple way so that possible  signatures of extra dimensions  in collider Physics can be estimated through various graviton KK modes~\cite{KK}-\cite{KK3}.
So far, we haven't seen any signal of warped extra dimensions in the current runs of LHC up to a few TeV and gravitational wave data could not rule out tiny warped dimensions as well~\cite{Pardo:2018ipy}.\\

In this work, we wish to look for the signatures of the Randall Sundrum 5D Braneworld Model in low-energy tabletop experiments  by studying the nature of quantum gravity-mediated entanglement between two masses in the presence of this warped extra dimension.
We will assume that the experiment is taking place in four dimensions, e.g. the state preparation and the creation of macroscopic Schr\"odinger Cat state is on our brane, while gravity can propagate in all the five dimensions, see Fig.~\ref{fig:my_label}. We wish to know how the entanglement at short distances manifests, especially when the distance between the two quantum systems is smaller than the $\rm AdS_5$ radius, i.e. we can probe the fifth dimension. In the infrared (IR), we will recover the results of four dimensions, and also the entanglement~\cite{Barker:2022mdz}. 

First, we shall briefly introduce the 5-D Randall Sundrum(RS) braneworld models~\cite{RandallSundrum1,RandallSundrum2}, and eventually will work in the backdrop of the RS single brane model(RS-2).
~We will consider the case when the scale of warping is lowered in comparison to \cite{RandallSundrum2}.This is well motivated from a phenomenological perspective since the other features of RS-2 model such as the continuum of KK gravitons do not lead to any observable collider signatures.
~Additional motivation comes from a class of Supergravity and String theories \cite{19} where it can be showed that the RS 3-brane cannot be identified with any D3 brane \cite{Kraus:1999it}~but is actually an effective geometry arising from a stack of negative tension branes stuck at the orbifold fixed point~\cite{Falkowski:2000er}. In this case, if we take a sufficiently large stack of these branes, then the warping scale can be lowered to the submilimeter regime and hence we should be able to detect the deviations from Newtonian gravity in the infrared tests for gravity~\cite{Chung}.~Since we have successfully tested Newtonian Potential up to $\rm52 \mu m$ ~\cite{Adelberger}, we can use this result to phenomenologically constraint the warping scale $k^{-1}\leq 52\mu m$.\\
~In section two, we perform a Kaluza-Klein (KK) decomposition and work out the tensor fluctuations about the RS-2 background and solve Einstein's equations. The resultant equations are analogous to quantum mechanical Schr\"odinger equations with the potential famously dubbed as the "volcano potential". This potential supports a single-bound state, which is the massless graviton and we recover the 4-D gravity on the brane. Additionally, we also obtain a continuum of massive KK graviton modes. Since we are interested to study the interaction between two matter particles via the exchange of virtual gravitons, we describe the structures of graviton propagators in 4-D when the underlying geometry is 5-D RS. The gravity on the brane is now  mediated by both massless and massive gravitons and we are interested to see how these massive modes can entangle masses on the 3-brane (our 4-D universe) in the infrared. We obtain the low energy limit of the tree-level scattering amplitude whose Fourier transform will yield the Newtonian Potential. In section IV, we then describe the setup for the matter system- two quantum harmonic oscillators carrying momentum along the x-direction which will get entangled due to the quantum gravitational interaction. In section V, we evaluate the Newtonian potential between two masses on the brane. We choose concurrence~\cite{Wootters,Hill,Balasubramanian} as the entanglement witness, and finally, in section VI, we calculate the concurrence in this setup and find that the concurrence now depends upon the radius of $\rm AdS_5$. We will show that for the harmonic oscillator case, the Gaussian wavepacket entangles but the concurrence is very tiny. To enhance the concurrence, we show that a spatial quantum superposition of masses is required, e.g. non-Gaussian state, or Schr\"odinger cat state.
We end with a discussion of our results.\\

Throughout the paper, the metric convention is mostly negative, and Greek indices $\alpha,\beta$ run from 0,1,2,3 and Latin indices a,b,c...denote 5D spacetime and run from 0,1,2,3,5 where 5 denotes the coordinate for extra-dimension.


\section{Brief Review of Randall Sundrum  Model}
\label{rs_review}
The RS1 Model \cite{RandallSundrum1} is a 5-D warped solution of Einstein's equations with two 3-branes of positive tension (Hidden brane) and negative tension(Visible brane) respectively embedded in the $AdS_{5}$ bulk. The extra-dimension is subject to $S_{1}/Z_2$ compactification with the fixed points identified. The RS action (in natural units) is:
\be
S=-\int{ d^5x \sqrt{-g}(M^3R-\Lambda)}+\int {d^4x\sqrt{-g_i}V_{i}}
\ee
where $\rm M$ is the 5-D Planck Scale,
~R is the 5-D Ricci Scalar, $\Lambda$ is the bulk cosmological constant,  $V_i$ is the tension of the ith brane(i=hid(vis)) and $\eta_{\mu\nu}=(+,-,-,-)$is the 4D metric. All Standard Model fields are confined on the visible brane and gravity alone can propagate in the warped extra-dimension.\\

The solution of Einstein's equations of motion gives us the metric:
 \be ds^2=e^{-2k |y|}\eta_{\mu\nu}dx^{\mu}dx^{\nu}-dy^2
 \label{RS metric}
 \ee
 supplemented by a negative bulk cosmological constant(the bulk is $\rm AdS_5$) $\Lambda=-12M^3k^2$ and brane tensions $V_{hid}=-V_{vis}=12M^3k^2$, where $k^{-1}$ is the radius of $AdS_{5}$. The extra-dimensional coordinate $0\leq y\leq\pi r_c$ and $r_c$ fixes the size of the extra-dimension.\\
 The 4-D Planck Scale $\MPl \approx \rm10^{19}GeV$ can be generated from $M$ on the visible brane according to the relation\cite{RandallSundrum1}:
 \be
 \MPl^2=\frac{M^3}{k}\LT 1-e^{-2k r_c \pi}\RT
 \label{mpl}
 \ee
 The above describes the two-brane setup. \\
 
 In \cite{RandallSundrum2}, it was proposed that it is possible to have a single 3-brane embedded in infinitely large warped extra-dimension by taking the brane at $y=\pi r_c$ to $\infty$.  In this case, the Planck/hidden brane of \cite{RandallSundrum1} becomes the visible brane, and this scenario(RS-2) describes an alternative to the standard KK compactification. The curved background supports the bound state of the five-dimensional massless graviton (m=0), thus reproducing the 4-D gravity on the 3-brane with modifications coming from the continuum of gapless massive KK graviton modes. We shall work in the backdrop of RS-2 and in our approach, we treat $k$ and $M$ as model parameters and constrain them from a phenomenological perspective.


\section{Non-relativistic Scattering in RS-2 Model}
\label{rs_grav}
We are interested to study the corrections to Newton's Law on the 3-brane when the underlying geometry is $\rm AdS_5$. Since the Newtonian Potential is the low energy limit of the tree-level scattering diagram of off-shell graviton exchange between two masses, we need to find the structure of the graviton propagator in this scenario. 

To  study the nature of gravity in this model, we will need to perform a KK reduction of the graviton in the ${\rm AdS}_5$ background. Due to compactification, we expect to see a graviton zero mode, a vector zero mode and a scalar zero mode that make up the five degrees of freedom in the 5-D graviton. At the massive level, we expect to see a tower of massive 4-D graviton modes which also makes up for five degrees of freedom. At the zero mode level of \ref{RS metric}, there would be a massless graviton and a massless scalar field (modulus field)(vector fields are ruled out due to $Z_2$ symmetry). For our case, we are interested in the tensor fluctuations, so we can set the scalar fluctuations to zero with a suitable gauge choice described below.
 \be 
 ds^2=A(y)^2\LF \eta_{\mu\nu}+h_{\mu\nu}\RF dx^{\mu}dx^{\nu}-dy^2\,,
 \label{fluctuations}
 \ee
where $A(y)=e^{-k |y|}$.The detailed KK decomposition of the graviton modes has been worked out  in \cite{TASI,RandallSundrum2,Callin} and we will only sketch the main ideas here for the sake of completion. 

We will be working in a gauge $h_{\mu}^{\mu}=0=\partial_\mu h^{\mu\nu}$ ($h_{55}$ and $h_{\rm a0}$ are also zero, thus reducing the independent degrees of freedom to two).Perform a coordinate transformation, $y\to z(y)$ where $z= sgn(y) \frac{1}{k}\LF e^{k |y|}-1\RF$. To perform a KK reduction down to four dimensions, we will separate the variables $h_{\mu\nu}(x,y)=\Tilde{h}_{\mu\nu}(x)\Tilde{\Phi}(y)$, where $\Tilde{\Phi}(y)=A^{-\frac{3}{2}}\Phi(z)$. Finally, we will require that $\Tilde{h}_{\mu\nu}(x)$ be a four-dimensional mass eigenstate mode $\Box \Tilde{h}_{\mu\nu}=m^2 \Tilde{h}_{\mu\nu}$  where $\Box=\eta^{\mu\nu}\partial_\mu \partial_\nu$ and m is the four-dimensional mass of the KK excitation. Hence $\Tilde{h}_{\mu\nu}(x)=e^{i p.x}$, where, $p^2=m^2$.Finally, the equation of motion for the KK modes can be recast in a form analogous to that of Schr\"odinger equation:
\be
  \left[ -\partial_z^2 + V(z) \right] \Phi_n(z)= m_n^2 \Phi_n(z)
  \label{Schrodinger}
\ee
where "n" labels the eigenstates and the effective potential $V(z)$ (volcano potential) is given by
\be
  V(z)= \frac{15k^2}{4\left(1 + k|z|\right)^2} - 3k \delta(z)
  \label{qm}
\ee
The delta function supports a single normalizable bound state which will be the 4-D massless graviton, and therefore we recover 4-D gravity on the brane. This result is consistent since we did not break the Poincare invariance in 4-D. Since the potential falls off to zero at infinity, we will also have continuum modes. Since the height of the potential near the origin is $\sim k^2$, the modes with $m^2<k^2$ will have suppressed wave functions, while those with $m^2>k^2$ will sail over the potential and hence un-suppressed near the origin.The solution to \ref{Schrodinger} is given in terms of Bessel Functions $J_n(x), Y_n(x)$ of order one and two:
\ba
 \Phi_n(z)=&& N_m \sqrt{1+k|z|}\LF Y_2[\tfrac{m_n}{k}(1+k|z|)]\Rd\non 
  &&\Ld-\frac{Y_1(\frac{m_n}{k})}{J_1(\frac{m_n}{k})}J_2[\tfrac{m_n}{k}(1+k|z|)]\RF
  \label{eq:u_solution}
\ea
where $N_m$ is the normalization constant.
The solution for zero modes is given by:
\be
  \Phi_0(z) = N_0 (1+k|z|)^{-3/2}.
  \label{eq:u_solution_zero}
\ee
The normalization constants $N_0$ and $N_m$ can be found by introducing a regulator brane at $z_r$ and then taking $z_r \to \infty$. Using delta function normalization: \cite{Callin}
\ba
  \int_{-\infty}^\infty |\Phi(0,z)|^2 dz =&& 1  \non
  \int_{-\infty}^\infty \Phi(m,z)^* \Phi(m',z) dz =&& \delta(m-m') 
   \label{eq:u_norm}
\ea
Consequently, it can be shown that the masses of graviton KK modes are quantized in the units of $\pi/z_r$.
\be
m_n\simeq \frac{n\pi}{z_r}\label{grav_mass}
\ee
where n=1,2,...When $z_r\to \infty$, we obtain a gapless continuum of massive modes. From now on, we can drop the index n in $m_n$. 
Finally, we can show
\be 
N_0=\sqrt{k}
\ee
and 
\be
  N_m^2 = \frac{\pi m}{2k z_r} \left[
    1 + \frac{Y_1^2(\frac{m}{k})}{J_1^2(\frac{m}{k})}
  \right]^{-1}
   \label{massive:norm}
\ee
\begin{figure}[h]
    \centering
    \includegraphics[width=200pt, max width=\textwidth]{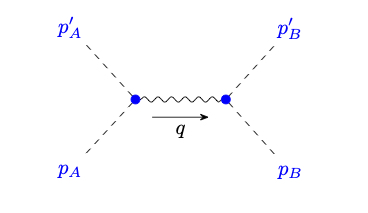}
    \caption{Tree-level scattering diagram of one graviton exchange.}
    \label{fig:scatterin}
\end{figure}
\FloatBarrier

Now that we have laid out the solution of \ref{Schrodinger} and found the behavior of massless and massive modes, we want to see these modes mediate interactions on the 3-brane. Therefore, we need to know the 5-D graviton propagator and the matter-graviton interaction term.
 Consider the tree-level scattering diagram shown in Fig.~\ref{fig:scatterin}, where q is the momentum of the off-shell graviton and 
$p_A$, $p_B$ are the momenta of the incoming spin-0 particles and $p_A^{\prime}$, $p_B{\prime}$ that of the outgoing particles. The off-diagonal quantum stress-energy tensor for spin-0 particles is \cite{Scadron}
\ba
   \langle p^\prime | T_{\mu\nu}(q)|p\rangle = \frac{1}{\sqrt{4 E E^\prime}}(p_{\mu}^\prime p_{\nu} + p_{\nu}^\prime p_{\mu}-\eta_{\mu\nu}(p^\prime \cdot p-m^2)\non
   \label{QEMT}
\ea
using the normalisation  $ \langle p |p'\rangle= 2 E (2\pi)^3\delta^3(p-p')$.~From now on, we will write $ \langle p^\prime | T_{\mu\nu}(q)|p\rangle$ as $T_{\mu\nu}(q)$ for brevity.
The propagator for a 5-D massless graviton $\hat{h}_{\rm{ab}}$ can be written as (assuming flat dimensions, we will later see that this is justified in our case) \cite{Giudice}:
\ba
  \mD^{(5)}_{\rm{abmn}}(x,y;x',y') \eqequiv
    \bra{0} \mathcal{T} \LF \hat{h}_{\rm ab}(x,y) \hat{h}_{\rm mn}(x',y')\RF \ket{0}
    \non
  & & \label{eq:5dpropagator}
\ea
where $\cal{T}$ denotes time ordering.Taking $y=y'=0$ (since both the particles are on the 3-brane and $y/z=0$ is the location of the 3-brane along the extra-dimension)
\ba
  \mD^{(5)}_{\rm{abmn}}(x,0;x',0) =  \int \frac{d^5 q}{(2\pi)^5}
    \frac{\mP_{\rm{abmn}}(q)}{q^2 +i\epsilon}
    e^{-iq \cdot (x-x')}
    \non
  & & \label{5dpropagator}
\ea
where
\ba
 \mP_{\rm{abmn}}(q) = \frac{1}{2}\LF
    \eta_{\rm am}\eta_{\rm bn} + \eta_{\rm an}\eta_{\rm bm} -\tth
    \eta_{\rm ab}\eta_{\rm mn}
  \RF
  \label{(eq:polarization_tensor5d)}
\ea
We have thus far shown that we can perform a KK reduction of 5-D graviton down to four dimensions. We should therefore be able to express the 5-D massless graviton propagator in terms of the 4-D propagator.\cite{Callin}.

The picture of a massless graviton propagating in D dimensions and the picture of massive KK gravitons propagating in 4 dimensions are equivalent, and from now on, we will use the former description in our discussion. Using the gauge conditions described before, we obtain,
\ba
&& \mD^{(5)}_{\mu\nu\alpha\beta}(x,0; x',0) =|\Phi(0,0)|^2 \mD^{(4,m=0)}_{\mu\nu\alpha\beta}(x,x')\non
    &&\hspace{20mm}+\sum_{m>0}^\infty |\Phi(m,0)|^2 \mD^{(4,m>0)}_{\mu\nu\alpha\beta}(x,x')
  \label{eq:5dpropagator_brane}
\ea
where $D^{(4,m=0)}_{\mu\nu\alpha\beta}(x, x')$ and $D^{(4,m>0)}_{\mu\nu\alpha\beta}(x, x')$ are the propagators of massless and massive 4-D spin-2 gravitons respectively.
\be
  \mD^{(4,m)}_{\mu\nu\alpha\beta}(x, x') =
    \int \frac{d^4 q}{(2\pi)^4}
    \frac{\mP^{(m)}_{\mu\nu\alpha\beta}(q)}{q^2 - m^2 + i\epsilon}
    e^{-iq \cdot (x-x')}
  \label{eq:4dpropagator}
\ee
For $m=0$,
\be
  \mP^{(m=0)}_{\mu\nu\alpha\beta}(q) = \frac{1}{2}\LF
    \eta_{\mu\alpha}\eta_{\nu\beta} + \eta_{\mu\beta}\eta_{\nu\alpha} -
    \eta_{\mu\nu}\eta_{\alpha\beta}
  \RF
  \label{(eq:polarization_tensor1)}
\ee
The polarization tensor for $m>0$ can be obtained by following the procedure of \cite{Giudice}, or from Fierz-Pauli Theory \cite{FP}. (The polarisation tensor for the massive gravity doesn't have a $m\to 0$ limit, and such limit can only be taken at the level of Lagrangian, which leads to the famous vDVZ discontinuity \cite{Z,vDV}. See \cite{MassiveGrav} for a detailed review).
\ba
  & & \hspace{2mm} \mP^{(m>0)}_{\mu\nu\alpha\beta}(q) =
    \frac{1}{2}\LF
      \eta_{\mu\alpha}\eta_{\nu\beta} + \eta_{\mu\beta}\eta_{\nu\alpha} -
      \eta_{\mu\nu}\eta_{\alpha\beta}
    \RF \non
  & & \hspace{3mm} -
    \frac{1}{2m^2} \LF
      \eta_{\mu\alpha} q_\nu q_\beta + \eta_{\mu\beta} q_\nu q_\alpha +
      \eta_{\nu\alpha} q_\mu q_\beta + \eta_{\nu\beta} q_\mu q_\alpha
    \RF \non
  & & \hspace{3mm} + \frac{1}{6}
    \LF \eta_{\mu\nu} + \frac{2}{m^2} q_\mu q_\nu \RF
    \LF \eta_{\alpha\beta} + \frac{2}{m^2} q_\alpha q_\beta \RF.
  \label{(eq:polarization_tensor2)}
\ea
Finally, we will need to consider the interaction with matter degrees of freedom. The graviton-matter interaction term in $D=5$  dimensions, is given by:
\be
\mathcal{L}_{int}=-\2 \rm T_{\rm  ab}h^{\rm ab}
\label{eq:int}
\ee
Since we assume matter to be confined on a 3-brane, the graviton-matter interaction term will not have the 55 component.Therefore,using $\rm M^3\sim k \MPl^2$ and $\rm h^{\rm ab}\to \rm M^{-3/2}h^{\rm ab}$, the required vertex is:\\
\begin{figure}[ht]
    \centering
    \includegraphics[width=200pt,max width=\textwidth]{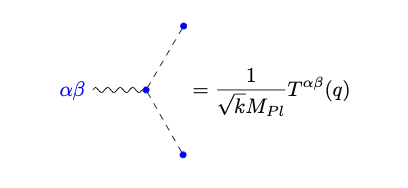}
    \label{vertex}
\end{figure}
\FloatBarrier
It is well known that the Newtonian Potential is the Fourier transform of low energy limit  (i.e. $q^0\to 0$) of the tree-level scattering amplitude \ref{fig:scatterin}, see \cite{Donoghue,Scadron}. There are various ways we can quantize graviton, we can either use Gupta formalism~\cite{Gupta,Gupta-1}, or impose gauge fixing and ghost degrees of freedom, see~\cite{Donoghue}. In either case, we can compute the change the gravitational energy due to the exchange of graviton~\cite{Marshman:2019sne,QGEM}. We are interested in this IR limit to study the nature of graviton-mediated quantum entanglement between the two particles on the 3-brane. Using \ref{eq:5dpropagator_brane}-\ref{eq:int}, it can be shown that \cite{Giudice,Callin}:
\be
  V(\mathbf{q}) = \lim_{q^0 \to 0} 
   \sum_{m=0}^\infty |\Phi(m,0)|^2 \frac{1}{k\MPl^2}
    \frac{T_1^{\mu\nu}(q) P^{(m)}_{\mu\nu\alpha\beta}(q) T_2^{\alpha\beta}(q)}
      {|q^2 - m^2|}\,
  \label{eq:NP}
\ee
Having determined the nature of the gravitational interaction, we want to lay out the setup for the matter sources which will  get entangled from a pure state via this quantum-gravitational interaction.


\section{SIMPLE SETUP FOR MATTER SYSTEM}\label{setup:matter}

For simplicity, we will consider two quantum harmonic oscillators in their respective traps separated by a distance $d$. The Hamiltonian is:
\be
\widehat{\Hc}_{matter}=\frac{\pha^2}{2m}+\frac{\phb^2}{2m}+\2 m \omega_m^2 \delta \xha^2 +\2 m \omega_m^2 \delta \xhb^2
\label{hamiltonian}
\ee 
where $\pha$,$\phb$ are the conjugate momenta, $\omega_m$ is the frequency of the oscillators and $\delta \xha$,$\delta \xhb$ are the quantum fluctuations around their mean values.
\ba
\xha=-\frac{d}{2}+\delta \xha, \hspace{8mm} \xhb=\frac{d}{2}+\delta \xhb
\label{displacements}
\ea
The mode operators for the harmonic oscillator systems are given by:
\ba
    \delta\xha = \sqrt{\frac{\hbar}{2 m \omega_m}}(a + a^\dagger) \, , \, \,
     \delta\xhb = \sqrt{\frac{\hbar}{2 m \omega_m}}(b + b^\dagger) \, , \, \,
     \label{xmode}
     \ea
     \ba
    \pha = i \sqrt{\frac{\hbar m \omega_m}{2}}(a - a^\dagger) \, , \, \, 
     \phb = i \sqrt{\frac{\hbar m \omega_m}{2}}(b - b^\dagger) \, 
     \label{pmode}
\ea
with the operators satisfying the usual canonical commutation relations.
Thus the Hamiltonian can be written as:
\be
    \widehat{\Hc}_{matter} = \hbar \omega_m \hat{a}^\dagger \hat{a} + \hbar \omega_m \hat{b}^\dagger \hat{b} 
\ee


\section{ENTANGLEMENT WITNESS}\label{EW}

We will assume that the  quantum harmonic oscillators are initially in their ground states.
The ground state of this system can be written as:
\be
\ket{\Psi_{i}}=\ket{0_A}\ket{0_B}
\ee
where $\ket{0_A},\ket{0_B}$ denote the number states and the respective ground states of oscillators A and B. Introduce a gravitational interaction between the oscillators. This interaction results in the oscillators becoming coupled if the gravitational interaction is quantum in nature, see~\cite{QGEM}. By following the standard perturbation theory procedure, the perturbed state is given by:
\be
\ket{\Psi_f}=\frac{1}{\mathcal{N}}\sum_{n,N}C_{nN}\ket{n}\ket{N}
\ee
where $\mathcal{N}=\sum_{n,N}|C_{nN}^2|$ is the overall normalisation factor.
The coefficient of the unperturbed state $C_{00}=1$ from the above formula, and that of the perturbed state is given by:
\be
C_{nN}=\lambda\frac{\bra{n}\bra{N} \widehat{\Hc}_{int}\ket{0}\ket{0}}{2E_0-E_n-E_N}\,,
\label{excitedstate}
\ee
 where $\lambda$ quantifies the strength of the interaction. We have dropped labels A and B from the states for ease of notation. 
We reiterate that it \textit{iff} $ \widehat{\Hc}_{int}$ is a quantum operator, only then we will \ref{excitedstate} yield a non-trivial value if $ \Hc_{int}$ is a classical C-number, the coefficients $C_{nN}=0$ due to orthogonality of the states. Non-zero $C_{nN}$ denotes the entanglement between the two states. Here $ \widehat{\Hc}_{int}$ designates the quantum interaction or the quantum communication, and it is an operator-valued entity that is compatible with the LOQC theorem, see~\cite{QGEM}.

Since we deal with the bipartite system, it is sufficient to witness the entanglement with the help of concurrence, which is defined by~\cite{Wootters,Hill}:
\be
    \mathcal{C} \equiv \sqrt{2(1-\text{Tr}\LT\rho_A^2\RT)} 
    \label{concurrence}
\ee
where $\widehat{\rho}_A$ is the density matrix of A, computed by tracing out the B state from the full density matrix.
\be
\widehat{\rho}_A = \rm \sum_{N}\langle N\ket{\rm \psi_f}\bra{\rm \psi_f}N\rangle
\label{A}
\ee
Substituting \ref{A} in \ref{concurrence}, we finally obtain:
\be
\mathcal{C} \equiv \sqrt{2\LF1-\sum_{n,n^\prime,N,N^\prime}C_{n,N}C^{\mbox{*}}_{n^\prime N}C_{n^\prime,N^\prime}C^{\mbox{*}}_{n N^\prime}\RF} 
\ee
The larger the concurrence, the more strongly entangled the subsystems are, where a maximally entangled state gives the value $\sqrt{2}$ and an unentangled state gives the value $0$.


\section{Effective potential}

Let us consider the non-static case, where we take the simplest scenario when 
$ p^\prime=p,{p^\prime}^2=p^2=m^2 $ in \ref{QEMT}, where~
 $p_{\mu}=(E/c, -\boldsymbol{p})$(introducing factors of c), $E=\sqrt{\boldsymbol{p}^{2}{c}^{2}+m^{2}{c}^{4}}$~and~$ (\mu, \nu)=(0,1)$.~Let $\boldsymbol{r}_{A}=\left(x_{A}, 0,0\right)$~and $\boldsymbol{r}_{B}=\left(x_{B}, 0,0\right)$ denote the positions of the two matter systems in one spatial dimension, assumed to be in the $x$-direction. While evaluating the effective potential \ref{eq:NP},~
      the Polarisation tensors \ref{(eq:polarization_tensor1)} and \ref{(eq:polarization_tensor2)} should be used for the massless and the massive KK modes respectively.
      Separating out the massless and the massive mode contributions 
      , we obtain
      \footnote{The momentum-dependent scattering is a new computation to our knowledge in the context of the RS-2 scenario. Previous computations concentrated on static scattering diagrams, e.g. static contributions to the Newtonian potential, see~\cite{RandallSundrum2,Callin,Giddings,Chung}. }
   \be
   V(\textbf{q})=V(\textbf{q})^{(m=0)}+V(\textbf{q})^{(m>0)}
   \ee
where, after taking the Fourier transform using the results $1/(2\pi)^3\int e^{-i \textbf{q.r}}/\textbf{q}^2=1/4\pi r$ and\\ $1/(2\pi)^3\int e^{-i \textbf{q.r}}/|\textbf{q}^2+m^2|=e^{-m r}/4\pi r$, we obtain:
\ba
V(r)^{(m=0)}= &&-\frac{G}{c^4 r}\LT {E}_A {E}_B+ \frac{{E}_B}{{E}_A}p_A^2 c^2+\frac{{E}_A}{{E}_B}p_B^2 c^2 \Rd\non
\hspace{8mm}&&\Ld +\frac{p_A^2 p_B^2 c^4}{{E}_A {E}_B}-4 p_A p_B c^2\RT 
\label{mm:NP}
\ea
where $r=|\bm r_A-\bm r_B|$ , the coupling $\MPl^{-2}=8\pi G/c^4$ and $|\Phi(0,0)|^2=k$. 
Expanding in the powers of $\tfrac{1}{c^2}$, upto $\mathcal{O}\LF \tfrac{1}{c^4}\RF$
\ba
V(r)^{(m=0)}= &&-\frac{G m^2}{r}-\frac{G}{ 2 r c^2}\LT 3 p_A^2 +3 p_B^2 -8 p_A p_B \RT \non
&&-\frac{G}{ 8 r m^2 c^4}\LT 18 p_A^2 p_B^2 -5 p_A^4 -5 p_B^4 \RT
\label{mm1:NP}
\ea

Contribution from massive modes:
\be
V(r)^{(m>0)}=-\frac{G}{c^4 k r} \sum_{m>0} |\Phi(m,0)|^2  e^{-m r} \LT T_A^{\mu\nu}P^{(m>0)}_{\mu\nu\alpha\beta}T_B^{\alpha\beta}\RT
\label{massive}
\ee
\vs
Here, exponential suppression is a characteristic of forces mediated by massive particles.

While writing \ref{massive}, we have used the fact that $T_A^{\mu\nu}$and $T_B^{\alpha\beta}$ are conserved matter sources on the brane. Therefore, $T^{(i)\alpha\beta}_{;\beta}=0$ for both the oscillators individually for $i=A, B$. Note that there are momentum-dependent contributions in the massive propagator, see \ref{(eq:polarization_tensor2)}. Hence, these terms will not contribute once we impose the condition of the conservation of energy-momentum tensor, e.g.\hs $q_\alpha T^{\alpha\beta}(q)= i \partial_\alpha T^{\alpha\beta}=0$. (Here we have replaced the covariant derivative with a partial derivative since we are working in a linearised theory). Therefore, both $P^{(m=0)}_{\mu\nu\alpha\beta}$ and $P^{(m>0)}_{\mu\nu\alpha\beta}$ are constants and depend only on the combinations of $\eta_{\mu\nu}$ and not on $q_\alpha, q_\beta$.
%

We now turn to evaluate the contribution from the massive modes for \ref{QEMT}. The normalization factor $|\Phi(m,0)|^2$ can be evaluated using \ref{massive:norm} and the sum over m can be converted to an integral in the limit $z_r\to \infty$ (recall that we have introduced a regulator brane at the conformal distance $z_r$ (See \cite{Callin},\cite{RandallSundrum2}). Using \ref{grav_mass},
\be
 \sum_m f(m)\to \int_{0} ^\infty  f(m)\frac{z_r}{\pi} dm
\ee
 and the property 
\be J_n(x)Y_{n+1}(x)-Y_n(x)J_{n+1}(x)=-\frac{2}{\pi x}\,\ee \ref{massive} becomes,
 \ba
V(r)^{(m>0)}&&=-\frac{2G}{\pi^2 r}\int_{0} ^\infty \frac{dm}{m} \frac{e^{-mr}}{J_1^2(\tfrac{m}{k}) + Y_1^2(\tfrac{m}{k})}\non
&&\hspace{8mm}\LF T_A^{\mu\nu} P^{(m>0)}_{\mu\nu\alpha\beta} T_B^{\alpha\beta}\RF
    \label{static1}
    \ea
  We can divide the integral in \ref{static1} into two regimes: ${m}/{k}<<1$ (light modes) and ${m}/{k}>>1$(heavy modes), and obtain up to leading order:
\ba
&&\int_{0} ^\infty dm \frac{e^{-mr}}{J_1^2(\tfrac{m}{k}) + Y_1^2(\tfrac{m}{k})}\approx\non
&&\int_{0} ^k dm  e^{-mr} \frac{m^2\pi^2}{4k^2}+\int_{k} ^\infty dm e^{-mr} \frac{m\pi}{2k} 
\label{small}
\ea
where for $m/k <<1$, the Bessel function of second kind $Y_1$ dominates in the denominator and for $m/k>>1$,\\
\ba
J_{n}(x)\approx \sqrt{\frac{2}{\pi x}}cos\LF x-\frac{n\pi}{2}-\frac{\pi}{4}\RF\\
Y_{n}(x)\approx \sqrt{\frac{2}{\pi x}}sin \LF x-\frac{n\pi}{2}-\frac{\pi}{4} \RF.
\label{large}
\ea
In this limit, the modes asymptote to plane waves.

Therefore, the contribution from the continuum of massive modes is :
\ba
&&V(r)^{(m>0)}\approx-\frac{G}{r c^4}\LT\frac{8}{3\pi^2}\RT \LT \frac{\pi^2}{ 4k^2 r^2}\LF 1-e^{-k r} (k  r+1)\RF \Rd \non
&&\Ld+\frac{\pi}{2 k r}\LF e^{-k r}\RF \RT\LT {E}_A {E}_B+\2 \LF \frac{{E}_B}{{E}_A}p_A^2 c^2+\frac{{E}_A}{{E}_B}p_B^2 c^2 \RF \Rd \non
&&\Ld+\frac{p_A^2 p_B^2 c^4}{{E}_A {E}_B}-3 p_A p_B c^2\RT 
\label{NP6}
\ea

Evaluating \ref{NP6} for $kr<<1$ and $kr>>1$ and expanding  in the powers of $\tfrac{1}{c^2}$ up to $\mathcal{O}\LF \tfrac{1}{c^4}\RF$ , we finally obtain up to leading order:
\ba\label{Vr}
V(r)^{(m>0)}\approx &&f(r)\Biggl\{-\frac{G m^2}{r}-\frac{G}{r c^2}\LT  p_A^2 + p_B^2 -3 p_A p_B \RT \non
&&-\frac{G}{ 8 r {m}^2 c^4}\LT 14 p_A^2 p_B^2 -3 p_A^4 -3 p_B^4 \RT\Biggr\}
\label{NP6:expansion}
\ea
where,
\be
  f(r)= \left\{ \begin{array}{lc}
    \ds \frac{4}{3\pi k r}+\mathcal{O}(1) \, , & k r<<1 \, , \vs \\
    \ds \frac{2}{3\pi (k r)^2} \, , & k r>>1  \, ,
  \end{array} \right.
\ee
The overall potential will be obtained by adding the contributions of the massless and massives modes. The static part of \ref{mm1:NP} and \ref{NP6:expansion} is:
\be
  V_0(r)\approx \left\{ \begin{array}{lc}
    \ds-\frac{G m^2}{r}\LT 1+\frac{4}{3\pi k r}\RT \, , & k r<<1 \, , \vs \\
    \ds -\frac{G m^2}{r}\LT 1+\frac{2}{3\pi (k r)^2}\RT \, , & k r>>1  \, ,
  \end{array} \right.
  \label{Vstatic}
\ee
This result can also be arrived at, following the procedure of \cite{Chung}.
In the limit $k r<<1$, the second term dominates in \ref{Vstatic}, and gravity behaves as a 5-D field with the potential falling off as ${1}/{r^2}$. In this limit, i.e. $r<<k^{-1}$, the length scale is smaller than the $\rm AdS_5$ radius, and hence the spacetime looks almost flat, as is seen by the ${1}/{r^2}$ fall off of the potential.  On the other hand, when $k r>>1$, it is the massless mode that dominates, and 4-D gravity is recovered with the KK modes providing corrections over the ${1}/{r}$ potential. In fact, this correction is above and beyond what one would expect from a single extra dimension. This is due to the barrier of the analog quantum mechanical problem \ref{qm} used to find the KK modes that result in the amplitude suppression of these modes near the brane. A beautiful description can be found in \cite{Lykken:1999nb}.

  Before we move on to the next section, we should remind that the full potential we have obtained, $V(r)=V(r)^{(m=0)}+V(r)^{(m>0)}$ see (\ref{mm:NP} - \ref{NP6}), have operator-valued entities. Since these potentials are obtained by assuming that the gravity is quantum in nature, e.g. $r, p_{A}, p_{B}$ are all operator-valued entities and not C-numbers. This has already been discussed in\cite{QGEM}.


\section{COMPUTING CONCURRENCE}
Assume that the particles on the 3-brane are exchanging graviton via the $\rm AdS_5$ (we are working in the limit $kr<<1$). We are interested to study how the entanglement builds up using concurrence as the entanglement witness, following the procedure of \cite{QGEM}. Promote \ref{QEMT} to the quantum mechanical stress-energy tensor for the quantum harmonic oscillators as per \ref{pmode}(as per Weyl quantization and interpret all the expressions in symmetrized ordering) and now the results should be interpreted as per \ref{setup:matter}.
From \ref{mm:NP}and \ref{NP6:expansion}, we can extract the terms that will eventually give us the lowest quantum matter-matter interactions. We are interested in studying the effective matter Hamiltonian by integrating the graviton degrees of freedom.
\ba
&&\widehat{\Hc}^{(0)}_{AB}\approx -\frac{G m^2}{\widehat{r}}\LT 1+ \frac{4}{3\pi k\widehat{r}}+ \RT + \cdots \label{H0}\\
&&\widehat{\Hc}^{(1)}_{AB}\approx 4\frac{G \pha}{\widehat{r} c^2}\LT1 +\frac{1}{\pi k\widehat{r}}  \RT \phb+ \cdots  \label{H1}\\
&&\widehat{\Hc}^{(2)}_{AB}\approx -\frac{G \pha^2 }{m^2 c^4 \widehat{r}}\LT\frac{9}{4}+\frac{7}{3\pi k\widehat{r}}\RT \phb^2 +\cdots
 \label{H2}
\ea
where $\Hc_{AB}$ are the effective Hamiltonian terms involving the static part of the potential and the momentum-dependent parts of the potentials. Note that we are just capturing the leading order contributions here in the perturbation theory, hence $\cdots$ terms, we will be neglecting them. These higher-order terms will contribute to the concurrence, but then they will be more suppressed.

\subsection{Contributions from \texorpdfstring{$\widehat{\Hc}^{(0)}_{AB}$}{H0}}
\label{gaussian}
Here we are capturing the static contribution at the leading order and see that at distances below the $\rm AdS_5$ radius, the second term dominates the Hamiltonian,
\be
\widehat{\Hc}^{(0)}_{AB}=-\frac{G m^2}{|\xha-\xhb|}\LT 1+ \frac{4}{3\pi k |\xha-\xhb|}\RT
\ee
where $\hat{r}=|\xha-\xhb|$=$|d+\LF\delta \xha - \delta \xhb\RF|^2$ (using \ref{displacements})
Expanding around $\delta x=|\xha-\xhb|=0$: 
\ba
&&\widehat{\Hc}^{(0)}_{AB}=-G m^2\LT \frac{1}{d}- \frac{1}{d^2}\LF\delta \xha - \delta \xhb\RF+\frac{1}{d^3}\LF\delta \xha - \delta \xhb\RF^2 \Rd\non
&&\Ld-\frac{4}{3\pi k}\LF \frac{1}{d^2}-\frac{2}{d^3}\LF\delta \xha - \delta \xhb\RF+\frac{3 }{d^4}\LF\delta \xha - \delta \xhb\RF^2\RF\RT
\ea
The lowest order quantum matter-matter interaction term is now given by:
\be \label{H_int0}
\widehat{\Hc}^{(0)}_{int}\equiv \frac{2 G m^2}{d^3}\LF 1+ \frac{4}{\pi k d}\RF \delta \xha \delta \xhb
\ee
Using mode expansions for $\delta \xha $ and $\delta \xhb$ as per \ref{xmode}, the above equation becomes:
\be
\widehat{\Hc}^{(0)}_{int}\equiv \hbar \textbf{\cal{g}}\LF \hat{a}\hat{b}+ \hat{a}^\dagger\hat{b}^\dagger+\hat{a}^\dagger\hat{b}+\hat{a}\hat{b}^\dagger\RF
\label{H_int}
\ee
where we have defined:
\be
\textbf{\cal{g}} \equiv\frac{Gm}{d^3 \omega_m}\LF1+\frac{4}{\pi k d}\RF
\ee
The first term is the same as obtained in the case of massless 4-D graviton \cite{QGEM} and the second term in the bracket is the correction to the coupling when we are probing the $\rm AdS_5$. The oscillators become strongly coupled in the presence of warped extra-dimension

Following the results in \ref{EW}, substituting \ref{H_int} in \ref{excitedstate}, the coefficient of the unperturbed state $C_{00}=1$ and that of the excited state is:
\be
C_{11}=-\frac{\textbf{\cal{g}}}{2 \omega_m}
\ee
The final state, up to the first order in perturbation theory, simplifies to:
\be
\ket{\Psi_f}=\frac{1}{\sqrt{1+\LF \textbf{g}/2 \omega_m\RF^2}}\LT\ket{0}_A\ket{0}_B-\frac{\textbf{\cal{g}}}{2 \omega_m}\ket{1}_A\ket{1}_B\RT
\ee
which is an entangled state involving the ground state and the first excited state of the system of two harmonic oscillators.
Finally, we can compute concurrence \ref{concurrence}
\be
\mathcal{C} \simeq \sqrt{2}\frac{\textbf{\cal{g}}}{\omega_m}=\sqrt{2}\frac{Gm}{d^3 \omega_m^2}\LF 1+\frac{4}{\pi k d}\RF
\ee
(under the assumption that $\textbf{g}/\omega_m<<1$).
Since $k d<<1$, the second term dominates over the first one and we can see that concurrence falls off with the quartic power of  the separation between the two oscillators. Therefore, we again see that in the presence of a quantum gravitational interaction, an un-entangled system has evolved into an entangled system. 

There will be parameters for which we will satisfy $\sqrt{2} >{\cal C}>0$ for witnessing entanglement. In the limit when $m \rightarrow \infty$, the concurrence vanishes, similarly  $\omega\rightarrow 0$, means the oscillators are no longer trapped and basically free, and the concurrence vanishes.
\begin{figure}[ht]
    \centering
    \includegraphics[width=230pt, max width=\textwidth]{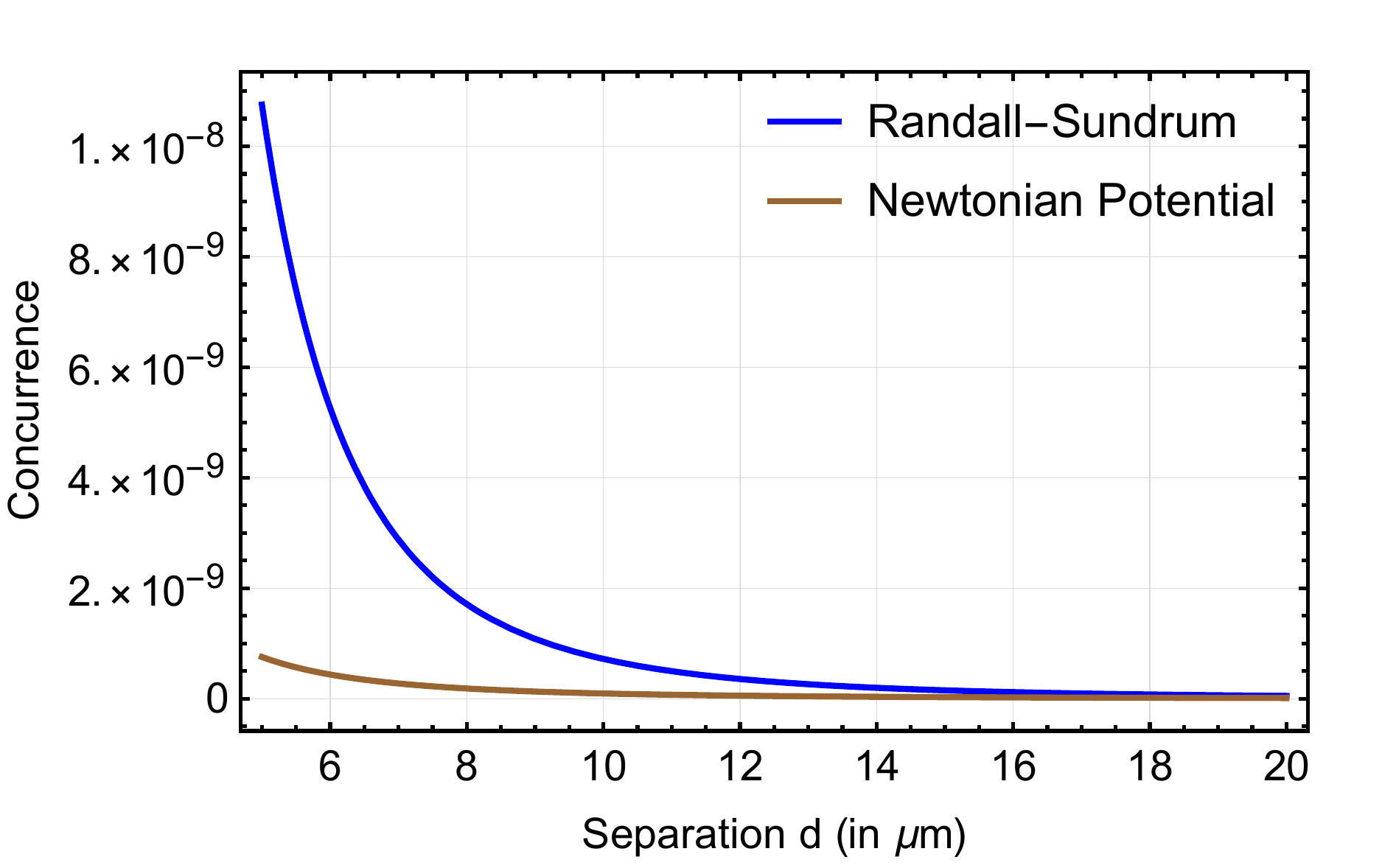}
    \caption{Concurrence as a function of the separation $d$ between the two harmonic oscillators,
    when $kd<<1$, where we have taken $k^{-1}\sim 52\mu m$, $m\sim 10^{-15}$kg and $\omega_m=1$Hz.}
    \label{fig:c1}
\end{figure}
\FloatBarrier
In the IR, gravity has been probed up to $52\mu m$, see \cite{Adelberger} and no deviations from Newtonian Potential have been observed. This sets the cutoff for new physics at below $52\mu m$. Therefore, to probe the effects of warped extra-dimensions in future table-top experiments, we can take $\rm AdS_5$ radius $k^{-1}\leq (52\mu m)$. By setting this value, we may consider similar parameter space as that of Ref.~\cite{Vankamp}, e.g. mass of order $m\sim 10^{-15}$kg object in a harmonic oscillator trap, with a frequency of $1$Hz ( feasible with diamagnetic trap~\cite{Hsu}).
If the underlying geometry is 5-D RS-2, then we should be able to witness 1-2 orders of magnitude enhancement in the concurrence as opposed to the predictions of 4-D Minkowski geometry, see Fig.~\ref{fig:c1}. Note that the concurrence for $kd \gg 1$ becomes similar to that of the 4-D scenario. Although the observed enhancement in concurrence is extremely small, so it will be extremely hard to witness this entanglement. One possibility could be to use entanglement tomography, by varying time or distance one can witness such a tiny signal~\cite{Barker:2022mdz,Bose:2022czr}. We can allow smaller trapping frequency, e.g. $\omega_m\sim 10^{-3}$Hz, the concurrence would decrease by $6$ orders of magnitude to ${\cal C}\sim 10^{-2}$. Such a low-frequency trap can be achieved by lowering the magnetic field gradient, or by lowering the value of Newton's constant, such as in a drop tower facility where the effective Newton's constant can be made small in a free-fall scenario. 

In fact, we can also increase the mass of the harmonic oscillators, however, by increasing the mass, the size of the object will considerably increase, which will be detrimental for us, since we are already within $d \sim {\cal O}(10) \mu m$ separation, so better the objects of interest must have masses below micron size, which is feasible for a diamond-like system. We will discuss this possibility when we discuss spatial superpositions instead of Gaussian harmonic oscillator states.

\subsection{Contributions from \texorpdfstring{$\widehat{\Hc}^{(1)}_{AB}$and $\widehat{\Hc}^{(2)}_{AB}$}{H1}}

We will now compute the concurrence for the lowest order quantum matter-matter interaction term in $\widehat{\Hc}^{(1)}_{AB}$ which will be dominated by the momentum operators $\pha$ and $\phb$:
\ba
\widehat{\Hc}^{(1)}_{int}= 4\frac{G \pha \phb}{ c^2 d}\LT 1+\frac{1}{\pi k d} \RT 
\ea
Writing $\widehat{\Hc}^{(1)}_{int}$ in terms of mode operators \ref{pmode},
\be
\widehat{H}^{1}_{int}\simeq \hbar \textbf{g}_1\LF \hat{a} - \hat{a}^\dagger\RF \LF \hat{b} - \hat{b}^\dagger\RF
\ee
where 
\ba
 \textbf{g}_1=&&\frac{2 G  m \omega_m}{c^2 d}  \LT 1+\frac{1}{\pi k d} \RT 
\label{H1mode}
\ea
Using \ref{H1mode} as the interaction term in \ref{excitedstate}, we can see that the only non-zero perturbation coefficient emerges from the term $\sim \hat{a}^\dagger\hat{b}^\dagger$:
\be
C_{11}=-\frac{\textbf{g}_1}{2 \omega_m}
\ee
The final state again is an entangled state involving the ground and the first excited states of the two harmonic oscillators.
\be
\ket{\Psi_f}=\frac{1}{\sqrt{1+\LF \textbf{g}_1/2 \omega_m\RF^2}}\LT\ket{0}_A\ket{0}_B-\frac{\textbf{g}_1}{2 \omega_m}\ket{1}_A\ket{1}_B\RT
\label{final1}
\ee
Using \ref{final1} in \ref{concurrence} for $\textbf{g}_1/\omega_m<<1$, we find the concurrence to be:
\ba
\mathcal{C} =&&\frac{2 \sqrt{2} G m }{c^2 d} \LT 1+\frac{1}{\pi k d} \RT 
\ea

The concurrence now linearly depends upon the mass of the quantum oscillators and inversely upon the $\rm AdS_5$ scale $k$.There is a $1/c^2$ suppression and hence the interaction $\widehat{\Hc}^{(1)}_{int}$ contribution is negligible and beyond the reach of any prospect of detectability.  For a similar set of parameters as that of the static case, we will get the concurrence of order ${\cal O}(10^{-36})$. Nevertheless, the trend would remain the same, the contribution from the massive graviton enhances the entanglement at short distances, below the warped radius $kd\ll 1$.


Similarly, the lowest order quantum matter-matter interaction terms in $\widehat{\Hc}^{(2)}_{AB}$ are:
\ba
&&\widehat{\Hc}^{(2)}_{AB}=-\frac{G \pha^2 \phb^2 }{m^2 c^4 d}\LT\frac{9}{4}+\frac{7}{3\pi k d}\RT
\ea
Writing $\widehat{\Hc}^{(2)}_{int}$ in terms of mode operators \ref{pmode},
\be
\widehat{\Hc}^{(2)}_{int}\sim -\hbar \textbf{g}_2 \LF \hat{a}^\dagger - \hat{a}\RF^2 \LF \hat{b}^\dagger - \hat{b}\RF^2
\ee
where 
\ba
&& \textbf{g}_2= \frac{G m \hbar \omega_m^2}{4 c^4 d}\LT\frac{9}{4}+\frac{7}{3\pi k d}\RT
\ea
Using \ref{H1mode} as the interaction term in \ref{excitedstate}, we can see that the only non-zero perturbation coefficient emerges from the term $\sim \LF\hat{a}^\dagger\RF^2\LF\hat{b}^\dagger\RF^2$:
\be
C_{22}=\frac{\textbf{g}_2}{2 \omega_m}
\ee
The final state is an entangled state involving the ground and the second excited state of the two harmonic oscillators.
\be
\ket{\Psi_f}=\frac{1}{\sqrt{1+\LF\textbf{g}_2/2 \omega_m\RF^2}}\LT\ket{0}_A\ket{0}_B+\frac{\textbf{g}_2}{2 \omega_m}\ket{2}_A\ket{2}_B\RT
\label{final2}
\ee
Using \ref{final2} in \ref{concurrence} for $\textbf{g}_2/\omega_m<<1$, we find the concurrence to be:
\ba
&&\mathcal{C}=\frac{\sqrt{2} G \hbar\omega_m }{ 4 c^4 d} \LT\frac{9}{4}+\frac{7}{3\pi k d} \RT
\ea
Concurrence depends linearly upon the reduced Planck's constant, mass of the oscillators and falls off with the $\rm AdS_5$ scale and square of the separation between the oscillators to the leading order. There is $1/c^4$ suppression in this case and experimentally, the effect of this term can only be detected for an experimentally impossible large frequency $\omega_m$.

\section{Spatial Superposition of two masses and concurrence}
\begin{figure}[ht]
	\centering
	 \includegraphics[width=230pt, max width=\textwidth]{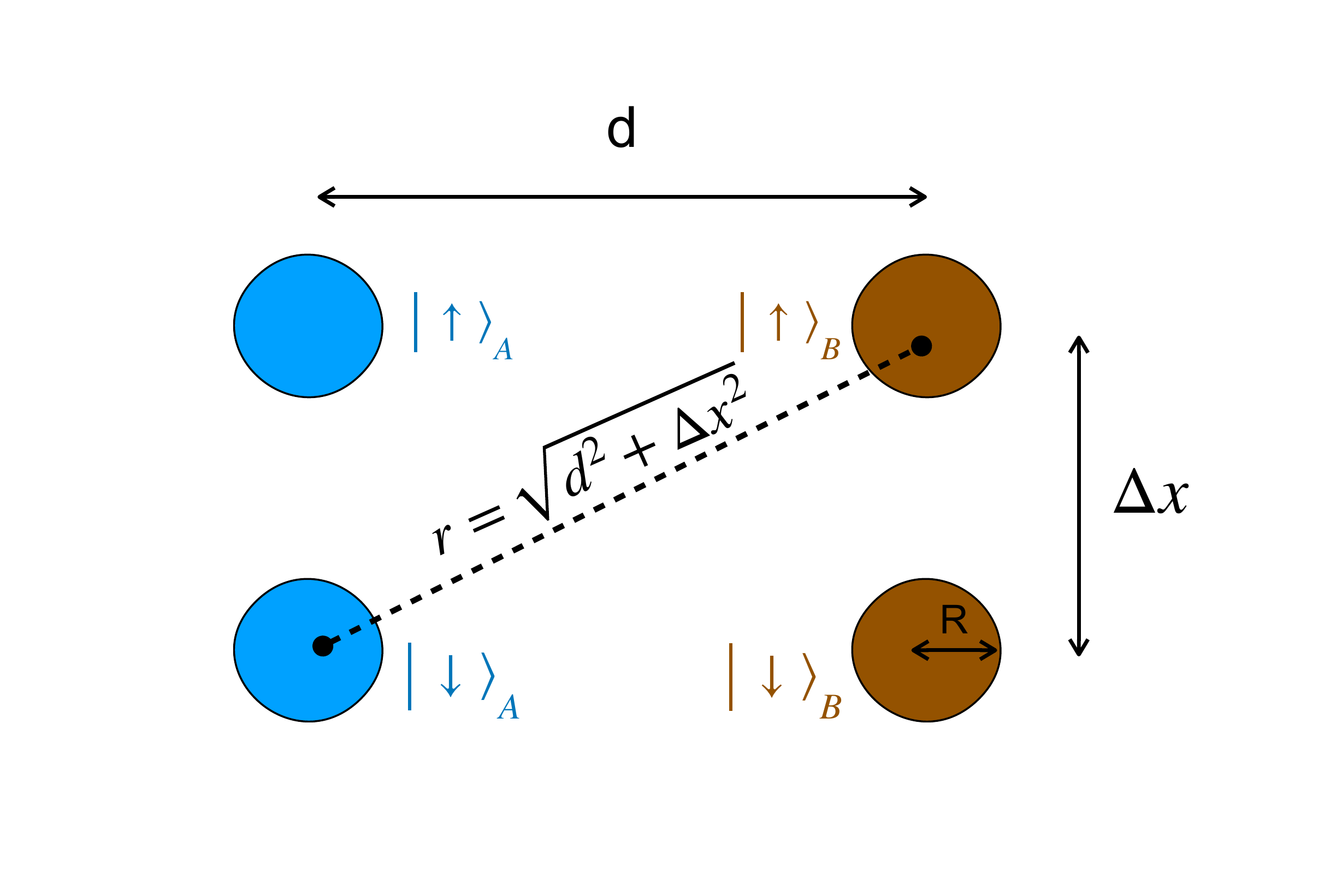}
	\caption{Configuration where the two spatial superpositions with the splitting $\Delta x$ are kept parallel to each other separated by distances $r$ and $d$ .The radius of the crystal is $R=\left({3m}/{4\pi\rho}\right)^{1/3}$, where we have taken $\rho$=3.5 g $cm^{-3}$ for a diamond-like system, where $R\ll \Delta x, d$, for masses $10^{-14}-10^{-15}$kg objects.  }
	\label{fig:system configuration}
\end{figure}

So far we have discussed the Gaussian state of the harmonic oscillator. We have seen that the concurrence is extremely tiny. However, this may change if we were to take a non-Gaussian state, such as spatial quantum superposition of masses. This can be achieved by the original QGEM protocol~\cite{Bose:2017nin}. Here, we will not delve into experimental challenges but we will consider a parallel setup discussed in \cite{Nguyen,Tilly:2021qef,Schut:2021svd,Barker:2022mdz}. Such spatial superpositions can be created in the Stern Gerlach setup with a nitrogen valence (NV) spin embedded in the crystal~\cite{Bose:2017nin}, see~\cite{Marshman:2021wyk,Yair}. Here we mainly discuss the concurrence in this setup. The masses are placed in a superposition of size $\Delta x$ in a parallel arrangement, shown in Fig. \ref{fig:system configuration}.

\FloatBarrier

The joint quantum states of the spins, assuming that the superposition is created at $t=0$ given by a separable state, see Refs.\cite{Bose:2017nin,Marshman:2019sne,Barker:2022mdz}
\be
\ket{\Psi(t=0)} =\frac{1}{2}\left[\left|\uparrow\uparrow\right\rangle + \left|\downarrow\downarrow\right\rangle + \left|\uparrow\downarrow\right\rangle + \left|\downarrow\uparrow\right\rangle\right] \nonumber
\ee
When the system interacts quantum gravitationally for time $\tau$, the wavefunction will evolve to an entangled state, given by~\cite{Bose:2017nin,Marshman:2019sne,Barker:2022mdz}:
\be
\ket{\Psi(t=\tau)} =\frac{1}{2}\left[\left|\uparrow\uparrow\right\rangle + \left|\downarrow\downarrow\right\rangle + e^{i\Delta\phi(d,r)}\left(\left|\uparrow\downarrow\right\rangle + \left|\downarrow\uparrow\right\rangle\right)\right] \nonumber
\ee
where the entanglement phase will be now given by 
$\phi(x)= \tau V_0 (x)/\hbar $. In the RS case, the effective 
potential is determined by \ref{H_int0}, such that
\be
\Delta\phi(d,r)=\frac{ G m^2 \tau}{\hbar}\LF \frac{1}{r}+\frac{4}{3\pi k r^2}-\frac{1}{d}-\frac{4}{3\pi k d^2}\RF
\ee
where d is the separation between the two masses and $r=\sqrt{d^2+\Delta x^2}$. For this setup, the density matrix for sub-system A can be obtained by tracing out the sub-system B from the full density matrix $\hat{\rho}$.
Therefore, 
\begin{align}
	\hat{\rho}_{A}=&\text{Tr}_{B}\left[\hat{\rho}\right] \nonumber\\
	=&  \frac{1}{2}\left[ {\begin{array}{cc}
		1 & \text{cos}\Delta\phi \\
		\text{cos}\Delta\phi & 1 \\
		\end{array} } \right]
\end{align}
and hence the concurrence $\cal{C}$ is given by:
\ba
 \mathcal{C}  &&\equiv\sqrt{2(1-\text{Tr}\LT\hat{\rho}_A^2\RT} \non
 &&= |\text{sin}\Delta\phi|
\ea

\begin{figure}[ht]
    \centering
    \includegraphics[width=230pt, max width=\textwidth]{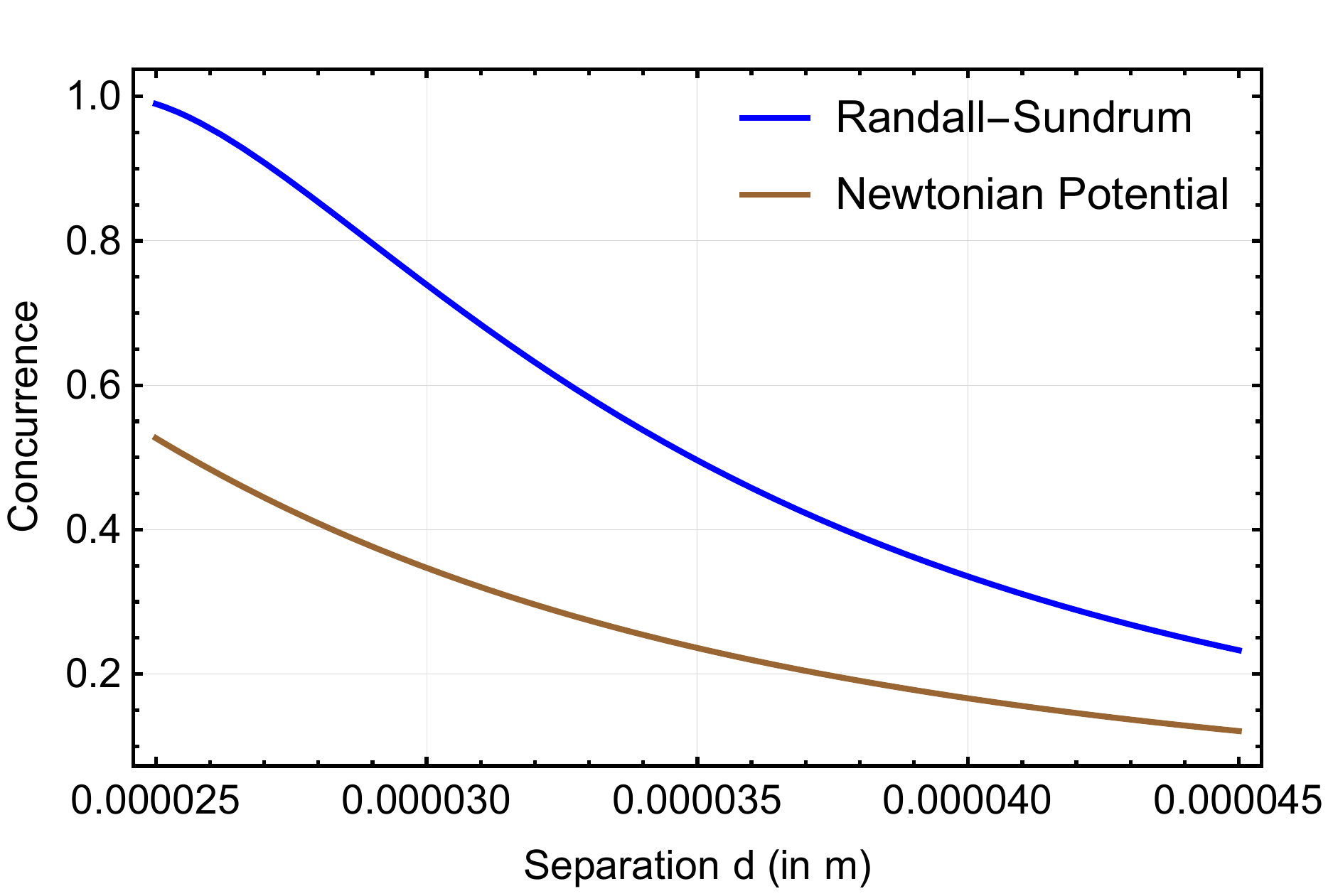}
    \caption{Concurrence as a function of the separation $d$ between the two spatial superpositions,
    when $kd<<1$, where we have taken $k^{-1}\sim 52\mu m$, 
    $m\sim 10^{-14}$kg, $\tau=1 s$ and $\Delta x=20\mu m$.}
    \label{fig:superposed}
\end{figure}
\FloatBarrier
For m=$10^{-14}kg,\Delta x=20\mu m, d\sim 40\mu m$ and $\tau$=$1$ second, the entanglement phase $\Delta\phi(d,r)=-0.341635$ and the concurrence $\cal{C}$=0.335028. We, therefore, obtain orders of magnitude enhancement in concurrence in comparison to the Gaussian state of the harmonic oscillator setup in \ref{gaussian}.

If gravity is fundamentally propagating in 4D Minkowski, then the entanglement phase $\Delta\phi(d,r)$ for this set of parameters would be $-0.16704$ and  the concurrence ${\cal C}=0.166265$. These results are encouraging, as they may provide us a possibility to probe the AdS physics if the scale is close to the vicinity of a few hundred microns. Here, of course, we have taken the AdS scale to be $k^{-1}\sim 52{\mu m}$($\implies \rm M\approx 10^7 GeV$ using \ref{mpl}), similar to the current constraints arising from any departure from Newtonian gravitational potential~\cite{Adelberger}. 

 We must note that the concurrence computed in this paper does not include the effect of decoherence. The most important channels for decoherence are indeed the electromagnetic channels, scattering due to ambient particles, discussed in these papers~\cite{Bose:2017nin,Vankamp,Tilly:2021qef,Schut:2021svd,Rijavec}. The decoherence rate due to the electromagnetic interaction dominates over any decoherence due to the gravitational interaction, see~\cite{Toros:2020krn,Danielson:2022tdw}. In particular, the creation of superposition will create tiny gravitational waves. However, it would depend on the details of the superposition. To excite gravitational waves, one will require the time variation in the quadrupole moment. Hence we will require either rotation of the diamond or an asymmetric superposition. The average energy of emitted gravitational waves will be dimensionally given by $\dot E\sim - GI^2\omega_m^6/c^2$, where $I$ is the moment of inertia and $\omega_m$ is the frequency of the trapping potential (assuming at the zeroth order the creation of superposition happens in a harmonic trap)~\cite{Gasperini}. Typically, for the mass range we are interested in and the smallness of the superposition size along with the frequency of the trapping potential (well within $10-100$Hz, means that the emission of gravitational waves is really tiny. One can also roughly estimate the emission rate of massless gravitons from the system, which we can dimensionally estimate to be 
$\gamma\sim t_{pl}^2\omega_{m}^3$~\cite{Toros:2020krn}. There is only one relevant time scale which is the trapping frequency of the superposition given by $\Omega_m$, and the $t_{pl}$ is the Planck time, e.g. $10^{-44}$ seconds. Again for $\omega_m\sim 10-100$ Hz, the emission rate is tiny, hence decoherence rate will be expected to be tiny as well. Of course, these numbers may change if we place the experiment near the black hole, as shown in~\cite{Danielson:2022tdw}. All these exercises are of academic interest and devoid of any experimental consequences.


\section{Summary}
 
In this paper, we have considered a very simple toy model of warped extra dimension, RS-2 scenario, where we probed the extra dimension via a protocol known as the quantum gravity-induced entanglement of masses (QGEM). We have obtained all our results relying only on the effective field theory of quantum gravity; the effective potential between the two masses was computed in a scattering theory, and the correction to the Hamiltonian has been computed up to the second order in perturbation theory. Both the wave function calculations and the correction to the Newtonian potential energy suggest that the quantum interaction between the graviton and the matter is crucial to obtain any entanglement, a classical description can not lead to entanglement. We computed the entanglement via concurrence and showed that the concurrence is always positive, although remains very tiny. We have shown that we will need large spatial splitting of the wavefunction, e.g. non-Gaussian state, to create a significant concurrence. Indeed, a large concurrence would also mean a significantly improved witness, provided the challenges of decoherence can be controlled appropriately~\cite{Bose:2017nin,Vankamp,Tilly:2021qef,Schut:2021svd,Rijavec,Gunnick,Toros:2020dbf}, and creating large superposition is possible in a laboratory. Note that the concurrence computed in this paper does not include the effect of decoherence. The most important channel for decoherence is indeed the electromagnetic channels, scattering due to ambient particles, discussed in these papers~\cite{Bose:2017nin,Vankamp,Tilly:2021qef,Schut:2021svd,Rijavec}. The decoherence rate due to the electromagnetic interaction dominates over any decoherence due to the gravitational interaction, see~\cite{Toros:2020krn,Danielson:2022tdw}. However, in this paper, we will not include these effects due to decoherence, and detecting the entanglement witness will require further analysis of the design of the experiment along with a detailed study of the decoherence rate. We will leave these for future studies.

Nevertheless, despite all these challenges our current study provides new ways of probing the physics beyond the Standard Model in the gravitational context. We have shown that for the non-Gaussian state the concurrence is significantly improved, for $m\sim 10^{-14}$kg quantum system, kept in a quantum superposition of $20{\rm \mu m}$, and separated by a distance $d<k^{-1}\sim 52{\mu m}$, the concurrence can be made order ${\cal C}\sim {\cal O}(0.1)$.

Indeed, it is a huge challenge to probe the parameter space of extra dimensions, which has already been constrained by the experiment to test the short-distance behavior of gravity~\cite{Adelberger}. Nevertheless, we believe that this modest approach taken in the current paper provides a quantum analog of the already existing tests of gravity, where we can also probe the quantum nature of both massless and massive graviton.\\

{\it Acknowledgments}:
SGE would like to thank Professor Soumitra Sengupta for his valuable insights. SGE is supported by IACS MS-Studentship.


\onecolumngrid

\end{document}